\documentclass{ws-mpla}

\newcommand{\be}{\begin{equation}}
\newcommand{\ee}{\end{equation}}
\newcommand{\ba}{\begin{eqnarray}}
\newcommand{\ea}{\end{eqnarray}}
\newcommand{\bc}{\begin{center}}
\newcommand{\ec}{\end{center}}

\begin{document}

\markboth{Diego F. Torres \& Eva Domingo-Santamar\'ia} {High
Energy emission from regions of star formation}

\catchline{}{}{}{}{}

\title{SOME COMMENTS ON THE HIGH ENERGY EMISSION FROM \\ REGIONS OF
STAR FORMATION BEYOND THE GALAXY\footnote{Solicited review article
for Modern Physics Letters A.}}

\author{\footnotesize DIEGO F. TORRES\footnote{
Present Address: Instituto de Ciencias del Espacio, Campus UAB,
Facultad de Ciencias Torre C-5, pares, 2a planta 08193 Barcelona,
Spain}}

\address{Lawrence Livermore National Laboratory, 7000 East Avenue,
L-413, Livermore, CA 94550, USA\\
dtorres@igpp.ucllnl.org}

\author{EVA DOMINGO-SANTAMARIA}

\address{Institut de F\'{\i}sica d'Altes Energies (IFAE),
Edifici C-n, Campus UAB, 08193 Bellaterra, Spain. \\
domingo@ifae.es }

\maketitle

\pub{Received (Day Month Year)}{Revised (Day Month Year)}

\begin{abstract}
Regions that currently are or have been subject to a strong
process of star formation are good candidates to be intense
$\gamma$-ray and neutrino emitters. They may even perhaps be sites
where ultra high energy cosmic rays are produced. Outside the
Galaxy, the more powerful sites of star formation are found within
very active galaxies such as starbursts (SGs) and Luminous or
Ultra-Luminous Infrared Galaxies (LIRGs or ULIRGs). Some general
characteristic of these objects are herein reviewed from the point
of view of their possible status as high energy emitters.

\end{abstract}

\ccode{PACS Nos.: }

\section{Introduction}

In this {\it Brief Review} we discuss some of the recent studies
on the high energy emission from regions of star formation located
in nearby galaxies, with emphasis on NGC 253.

\section{Diffuse $\gamma$-ray emission from galaxies}


The diffuse $\gamma$-ray emission observed from a galaxy basically
consists of three components: the truly diffuse emission from the
galactic interstellar medium (ISM) itself, the extragalactic
background, and the contribution of unresolved faint point-like
sources that belong to the galaxy or are beyond it but in the same
line-of-sight by chance. The galactic diffuse emission generated
in interactions with the ISM has a wide energy distribution and
normally dominates the other components. $\gamma$-rays are thus
mostly produced in energetic interactions of particles with the
interstellar gas and the radiation fields present in the galaxy.
The diffuse $\gamma$-ray emission {at high energies} mainly comes
from the interaction of high energy cosmic ray nucleons with gas
nuclei via neutral pion production. {Contributions from} energetic
cosmic ray electrons interacting with the existing photon fields
via inverse Compton scattering and with the matter field of the
galaxy via relativistic bremsstrahlung {are generally more
important below ~100 MeV. Since these} processes are dominant in
different parts of the energy domain, information about the
overall spectrum of the cosmic ray population can be extracted.

If $\gamma$-rays absorption in the galaxy is neglected, the
diffuse $\gamma$-ray flux can be estimated as the line-of-sight
integral over the emissivity of the ISM. The latter is essentially
the product of the cosmic ray spectrum, the density of the gas or
radiation field in the galaxy, { and the corresponding cross
section for a given process.} Therefore, $\gamma$-ray measurements
together with estimations of the gas content and photon field
densities provide a tool to determine the cosmic rays spectrum.
The interstellar hydrogen distribution (in its molecular form,
H$_2$, atomic, HI, or ionized,  HII) is derived from radio
surveys. It can be traced from the emission lines of molecules
that get radiatively or collisionally excited, by means of
corresponding calibrations or conversion factors. CO molecules are
generally the primary tracers of molecular hydrogen (e.g., Dame et
al. 2001). CO is a polar molecule with strong dipole rotational
emission at millimeter-wavelengths. H$_2$, though much more
abundant than CO, has only a weak quadrupole signature. Generally,
the conversion factor between CO luminosity and molecular mass is
estimated and constrained from measurements in our own Galaxy
where masses of individual molecular clouds can be independently
determined from cloud dynamics (e.g., Solomon et al. 1987; Young
\& Scoville 1991). Alternatively, assuming the spectrum of cosmic
rays of the Galaxy and a given $\gamma$-ray flux, it is possible
to obtain a calibration for the CO luminosity to estimate the
conversion factor (e.g., Bloemen et al. 1986). The atomic neutral
hydrogen content can be estimated trough the intensity of 21-cm
emission line. HCN, CS and HCO$^{+}$ are the most frequently
observed interstellar molecules after CO. Due to their higher
dipole moment
they require about two orders of magnitude higher gas densities
for collisional excitation than CO.
HCN is one of the most abundant high dipole moment molecules and
traces molecular gas at densities $n(H_2)\gtrsim 3 \times 10^4$
cm$^{-3}$, compared to densities of about $\gtrsim 500$ cm$^{-3}$
traced by CO. {Because of this fact, dense regions associated with
star formation sites are usually better traced at the higher HCN
frequencies.}

\begin{figure*}[t]
\centering
\includegraphics[width=0.62\textwidth]{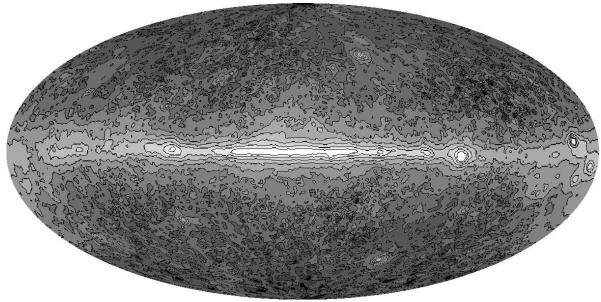}
\caption{Intensity of $\gamma$-rays ($>100$ MeV) observed by
EGRET. The broad, intense band near the equator is interstellar
diffuse emission from the Milky Way. The intensity scale ranges
from $1 \times 10^{-5}$ cm$^{-2}$ s$^{-1}$ sr$^{-1}$ to $5 \times
10^{-4}$ cm$^{-2}$ s$^{-1}$ sr$^{-1}$ in ten logarithmic steps.
The data is slightly smoothed by convolution with a gaussian of
FWHM 1.5$^\circ$. Courtesy: EGRET Collaboration.} \label{egretall}
\end{figure*}


\subsection{Diffuse $\gamma$-ray emission from the Galaxy}

Studies of the Galactic diffuse $\gamma$-ray emission provide
privileged insights into the generation processes of $\gamma$-rays
in galactic environments. The diffuse $\gamma$-ray continuum
emission is in fact the dominant feature of the $\gamma$-ray sky
of the Milky Way (see the EGRET $\gamma$-ray map above 100 MeV
presented in Figure \ref{egretall}), approximately amounting 90\%
of the high energy $\gamma$-ray luminosity ($\sim$1.3 $\times
10^6$ L$_\odot$, Strong, Moskalenko, \& Reimer 2000). This
emission, in the range of 50 KeV to 50 GeV, was systematically
studied by all the past hard X-ray/$\gamma$-ray satellites, from
SAS-2 and COS-B in the 70's and early 80's, to the OSSE, COMPTEL
and EGRET experiments onboard of the Compton Gamma-Ray Observatory
(CGRO), launched in 1990, {as well as, more recently, by INTEGRAL
(Lebrun et al. 2004)}. Hunter et al. (1997) present a review of
CGRO observations.

\subsection{Detection (and non-detection) of other local group galaxies}

\begin{figure*}[t]
\centering
\includegraphics[width=0.62\textwidth]{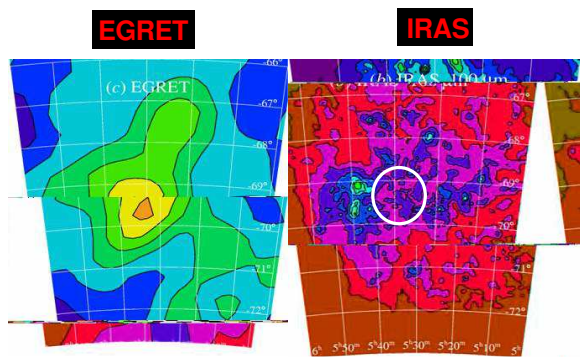}
\caption{LMC as seen by EGRET ($\gamma$-rays) and IRAS (infrared).
White circle indicates the position of 30 Doradus, a large
molecular cloud and intense star formation region. {Courtesy: Seth
Digel.}} \label{LMC}
\end{figure*}

To date, the Large Magellanic Cloud (LMC) is the only external
galaxy that has been detected in the light of its diffuse
$\gamma$-ray emission (Sreekumar et al. 1992). This fact is
explained by the isotropic flux dilution by distance. At 1 Mpc,
for example, the flux of the Milky Way would approximately be 2.5
$\times 10^{-8}$ photons cm$^{-2}$ s$^{-1}$ above 100 MeV, well
below the sensitivity achieved up to now by the $\gamma$-ray
missions in the relevant energy domain. Although normal galaxies,
or even galaxies with fairly intense star formation such as the
LMC, are quite numerous, their distances ought to make them very
faint $\gamma$-ray sources.

The LMC was detected by EGRET with a flux of (1.9$\pm$0.4)
$\times$ 10$^{-7}$ photons cm$^{-2}$ s$^{-1}$ above 100 MeV (see
Figure \ref{LMC}). Interestingly, as can be noticed in Figure
\ref{LMC}, the distribution of the diffuse $\gamma$-ray emission
from the LMC is consistent with the infrared IRAS map profile,
being the more intense $\gamma$-ray emission region in spatial
coincidence with the position of 30 Doradus, a particular region
of the LMC with large molecular clouds and extensive ongoing star
formation. {This correspondence is indeed the reason why the EGRET
team claimed the detection of the LMC.} This result had been
predicted by Fichtel et al. (1991) as the output of pion decay
resulting from the interaction between cosmic ray protons and
interstellar gas, assuming galactic dynamic balance between the
expansive pressures of the cosmic rays, magnetic fields and
kinematic motions, and the gravitational attraction of matter.

{It is instructive to show how to obtain the predicted flux and
what consequences it has in our understanding of cosmic ray
generation. One can consider that the electron spectrum is a
power-law $N(E) dE= KE^{-\gamma} dE$, with $N(E)$ being the number
of electrons per unit energy per unit volume, and $K$ the spectrum
normalization. The intensity of the synchrotron radiation in the
presence of random magnetic fields is \be I_\nu=1.35 \times
10^{-22} a(\gamma) L\, K\, B^{(\gamma+1)/2} \left(\frac{6.26
\times 10^{18}}{ \nu} \right)^{(\gamma-1)/2}
 {\rm erg\; cm}^{-2} \; {\rm s}^{-1} \; {\rm
sr}^{-1}\; {\rm Hz^{-1} }, \label{1} \ee (Ginzburg \& Syrovatskii
1964) where $\nu$ is the observing radio frequency in Hz,
$a(\gamma)$ is a numerical coefficient of order 0.1, $L$ is the
length over which the electrons and magnetic fields are present
and $B$ is the magnetic field strength. The normalization of the
spectrum is assumed proportional to $B^2$ both in the LMC and our
Galaxy, and the shape of the spectrum in the LMC is assumed the
same as that in the Milky Way. Then, if $K_0$ and $B_0$ are the
corresponding values of these parameters in our Galaxy, and $w(x)
K_0$ is the value in the LMC, $B=w(x)^{1/2}B_0$. Using this
expression in Eq. (\ref{1}) the scaling can be determined as \be
w(x)=\left( \frac{2.40I_\nu}{a(\gamma) L_{21} K_0}
\right)^{4/\gamma+5} B_0^{-2(\gamma+1)/(\gamma+5)} \left(
\frac{\nu}{6.26 \times 10^{18}}\right)^{2(\gamma+1)/(\gamma+5)},
\ee where $L_{21}=L/(3.09\times 10^{21}{\rm cm})$ is the distance
in kpc. Assuming best guesses for all parameters involved (see,
for instance, the Appendix of Fichtel et al. 1991), the electron
normalization can be determined. The additional assumption that
the electron-to-proton ratio is the same in the LMC as in the
Galaxy yields the proton spectrum. An estimation of the matter
column density then allows the $\gamma$-ray flux to be computed
as:
\begin{eqnarray} F(E>100 {\rm MeV}) \simeq \int d\Omega \left[ 2 \times
10^{-25} \times \frac{w(x)}{4\pi} \times \int dl (n_a +n_m)
\right]\; \frac 1{4\pi d^2}\hspace{1.5cm}\nonumber \\   {\rm
photons\; s}^{-1}\; {\rm cm}^{-2},
\end{eqnarray} with $d\Omega$ being the solid angle subtended by the emitting
region, $j_\gamma=2 \times 10^{-25} \times {w(x)}/{4\pi}$ photons
s$^{-1}$ sr$^{-1}$ H-atom$^{-1}$ being the $\gamma$-ray
production, and $ \int dl (n_a +n_m)$, with $n_a$ and $n_m$ the
atomic and molecular density, respectively, being the column
density. Note that the prediction allows different emission level
contours to be plotted, depending on the position in the galaxy.
However, in order to make a direct comparison with EGRET or any
other experiment, the predicted $\gamma$-ray intensity has to be
compared with the corresponding point-spread function. Although
the predicted intensity based on the dynamic balance is in good
agreement with the EGRET result, it is also in agreement with the
cosmic ray density being the same throughout the galaxy, as if,
for instance, the cosmic ray density were  universal in origin (as
proposed, for instance, by Brecher and Burbidge 1972).}

The Small Magellanic Cloud (SMC) was also observed by EGRET, but
no detection was found. An upper limit of 0.5 $\times$ 10$^{-7}$
photons cm$^{-2}$ s$^{-1}$ was set for the $\gamma$-ray emission
above 100 MeV (Sreekumar et al. 1993). If the cosmic ray density
in the SMC were as high as it is in our Galaxy, the flux in
$\gamma$-rays would be $\sim$ 2.4 $\times$ 10$^{-7}$ photons
cm$^{-2}$ s$^{-1}$, a level incompatible with the experimental
result. It was then the non-detection of the SMC what defined that
the distribution of cosmic rays is galactic in origin, local to
accelerators and thus enhanced in regions of star formation rather
than universal.

{Very recently, Pavlidou and Fields (2001) presented an
observability study for several of the local group galaxies,
assuming that the $\gamma$-ray flux above 100 MeV is represented
by \be  F(E>100 {\rm MeV})= 2.3 \times 10^{-8} f_G\left( \frac
{\Sigma}{10^4 {\rm M}_{\odot} {\rm kpc} ^{-2}}\right) {\rm photons
\, cm}^{-2} \, {\rm s}^{-1}, \ee with $f_g$ being the ratio
between the galaxy $G$ and the Milky Way supernova rates, and
$\Sigma$ the gas mass-to-distance squared ratio. This amounts to
the assumption that supernova remnants alone are the source of
cosmic rays, and that once produced, their propagation is
described by a leaky box model, with the additional supposition of
an equal time/length of escape to that of our Galaxy. This
approach is far simpler than that followed by Fichtel, Sreekumar
and coworkers when analyzing the LMC and SMC cases, and probably
not quite correct, especially for those galaxies which are
different from ours, like the SMC. For example, the Andromeda
galaxy M31, a case studied previously by \"Ozel and Berkhuijsen
(1987), would present a flux of $1 \times 10^{-8}$ photons
cm$^{-2}$ s$^{-1}$, consistent with the observational upper limit
set by Blom et al. (1999) using more recent EGRET data. This flux
could be detected by GLAST in the first 2 years of its all-sky
survey with 14$\sigma$ significance. If such is the case, it will
be possible to study the correlation between regions of higher
column density and higher $\gamma$-ray emission. It could even be
possible to observe effects of the magnetic torus (e.g., Beck et
al. 1996) and the star forming ring (e.g., Pagani et al. 1999), a
morphological feature analogous to the Milky Way's H$_2$ ring
extending in radius from 4 to 8 kpc (e.g., Bronfman et al. 1988),
which has been detected in $\gamma$-ray surveys (Stecker et al.
1975). Other results for Local Group Galaxies show that, unless
the assumptions are severely misrepresenting the physics, only M33
might have some chance of being detected by future instruments
(Digel et al. 2000).}

\section{Galaxies with higher star formation rate}

Galaxies where star formation is a powerful active process may be
able to compensate the dilution effect produced by their distance
to Earth. The large masses of dense interstellar gas and the
enhanced densities of supernova remnants and massive young stars
expected to be present in such galaxies suggest that they emit
$\gamma$-ray luminosities orders of magnitude greater than normal
galaxies. Such environments will typically emit a large amount of
infrared (IR) radiation, because abundant dust molecules absorb
the UV photons emitted by the numerous young massive stars and
remit them as IR radiation. Therefore, the infrared luminosity,
$L_{\rm IR}$, of a galaxy can (but not always) be an indication of
star formation taking place in it.

Luminous infrared galaxies (LIRGs) have been identified as a
class, generally selected  for emitting more energy in the IR band
($\sim$ 50 -- 500 $\mu$m) than in all other wavelengths combined.
The most luminous galaxies in the IR band constitute by themselves
a subclass of the more powerful galaxies ever known. LIRGs are
defined as galaxies with IR luminosities larger than 10$^{11}
L_{\odot}$. Those with $L_{\rm IR} >$ 10$^{12} L_{\odot}$ are
called ultraluminous infrared galaxies (ULIRGs). See Sanders \&
Mirabel (1996) for an extensive review about these objects.
LIRGs are the dominant population of extragalactic objects in the
local universe ($z<0.3$) at bolometric luminosities above $L >
10^{11}$ L$_\odot$, ULIRGs are in fact the most luminous local
objects.
Our current understanding of LIRGs and ULIRGs suggests that they
are recent galaxy mergers in which much of the gas of the
colliding objects, particularly that located at distances less
than $\sim 5$ kpc from each of the pre-merger nuclei, has fallen
into a common center (typically less than 1 kpc in extent),
triggering a huge starburst (e.g., Sanders et al. 1988, Melnick \&
Mirabel 1990). The size of the inner regions of ULIRGs, where most
of the gas is found, can be even as small as a few hundreds
parsecs; there, a large nuclear concentration of molecular gas is
found.
Supporting the idea that IR luminosities in LIRGs are mainly due
to starburst regions rather than to enshrouded AGNs it is the fact
that LIRGs not only  possess a large amount of molecular gas, but
a large fraction of it is at high density (e.g., Gao \& Solomon
2003a, 2003b). This  makes them prone to star formation, and thus
to have significant CR enhancements.
In addition, there is evidence for the existence of extreme
starbursts regions within LIRGs  (see, e.g., Downes \& Solomon
1998). These, larger than giant molecular clouds but with
densities found only in small cloud cores, appear to be the most
outstanding star-forming regions in the local universe (each
representing about 1000 times as many OB stars as 30 Doradus).
They are well traced by HCN emission, i.e., they represent a
substantial fraction of the whole HCN emission observed for the
whole galaxy (Downes \& Solomon 1998, Solomon et al. 1992). The CR
enhancement factor in these small but massive regions can well
exceed the average value for the galaxy.


To date, no LIRGs or ULIRGs, nor any other starburst galaxy has
been detected in $\gamma$-rays by EGRET, upper limits were imposed
for M82,  $F(E>100 {\rm MeV}) < 4.4 \times 10^{-8}$ photons
cm$^{-2}$ s$^{-1}$, and NGC~253, $F(E>100 {\rm MeV}) < 3.4 \times
10^{-8}$ photons cm$^{-2}$ s$^{-1}$ (Blom et al. 1999), the two
nearest starbursts. Similar constraints were found for many LIRGs
by means of a search in existing EGRET data for the fluxes of
likely $\gamma$-ray--bright LIRGs (Torres et. al. 2004, Cillis et
al. 2005).

\subsection{Should we detect them?}

Neglecting possible CR density gradients within the interstellar
medium of the galaxy, the hadronically-generated $\gamma$-ray
number luminosity (photons per unit of time) is given by:
\be I_{\gamma}(E_{\gamma})=\int n(r) q_{\gamma}(E_{\gamma}) dV =
\frac{M}{m_p} q_{\gamma} \,, \ee
where $ r$ represents the position within the interaction region
$V$, $M$ is the mass of gas, $m_p$ is the proton mass, $n$ is the
number density, and ${q_\gamma}$ is the $\gamma$-ray emissivity
(photons per unit of time per atom).
The $\gamma$-ray flux is then:
\be F(>100\,{\rm MeV})=\frac{I_\gamma(>100\,{\rm MeV})}{4\pi
{D_L}^2} \,, \ee
where $D_L$ is the luminosity distance in a Friedman
universe.\footnote{\label{dlfoot}The luminosity distance describes
the distance at which an astronomical body would lie based on its
observed luminosity. If a source of luminosity $L$ emits into a
non-expanding universe, the integral of the photon flux over a
sphere of given radius will be equal to the source luminosity. In
an expanding space-time, the photon wavelength is redshifted,
diluting the energy over the sphere by a factor of (1+$z$), where
$z$ is the redshift. Another factor (1+$z$) is introduced as the
emission rate of the photons from the source is time dilated with
respect to an observer due to the Doppler effect. Therefore, the
observed luminosity is attenuated by two factors: relativistic
redshift and the Doppler shift of emission, each of them
contributing a (1+$z$) attenuation. The luminosity distance can be
expressed in terms of the Hubble parameter, $H_0$, the
deceleration parameter, $q_0$, and the redshift: \be D_L =
\frac{c}{H_0 q_0^2} \left[ 1 - q_0 + q_0 z + (q_0-1) \sqrt{2 q_0 z
+ 1} \right]. \ee }
In an appropriate scaling, the $\gamma$-ray flux can be estimated
from:
\be F(>100\; {\rm MeV}) \sim 2.4 \times 10^{-9} \left(
\frac{M}{10^9 {\rm M}_\odot} \right) \left( \frac{D_L}{\rm Mpc}
\right)^{-2} k \;\;\; {\rm photons\; cm^{-2} s^{-1}} \,.
\label{Fglast}\ee
The previous estimation introduces $k$ as the enhancement factor
of $\gamma$-ray emissivity in the galaxy under study compared to
the local value near the Earth. If the slope of the CR spectrum at
the galaxy does not differ much from that existing near the Earth,
what clearly is a rough approximation, $k$ can be at the same time
an estimator of the enhancement of CR energy density: $k \equiv
{q_\gamma}/{q_{\gamma,\oplus}} \sim \omega / \omega_\oplus$, with
${q_{\gamma,\oplus}} = 2.4 \times 10^{-25}$ photons s$^{-1}$
H-atom$^{-1}$ being the $\gamma$-ray emissivity of the
interstellar medium at the Earth neighborhood, and $\omega_\oplus$
being the CR energy density near Earth. The numerical factor of
Equation (\ref{Fglast}) already takes into account the
$\gamma$-ray emissivity from electron bremsstrahlung (see, e.g.,
Pavlidou \& Fields 2001 and references therein).\footnote{Note
that $\gamma$-rays can also be produced by inverse Compton
interactions with the strong FIR field of the galaxy. However,
this contribution has been disgarded in favor of the hadronic
channel (between accelerated protons and diffuse material of
density $n$), which is a well justified approach above 100 MeV
(see below). We also disregard additional hadronic production of
high energy $\gamma$-rays with matter in the winds of stars (see
e.g. Romero \& Torres 2003, Torres et al. 2004b). } Note that $
F(>100\; {\rm MeV}) \sim 2.4 \times 10^{-9} \;{\rm photons\;
cm^{-2} s^{-1}} $ is approximately the GLAST satellite sensitivity
after 1 yr of all-sky survey.
Therefore, from Equation (\ref{Fglast}), having an estimation of
the mass gas content of the galaxy through the measured CO
luminosity, the minimum average value of $k$ for which the
$\gamma$-ray flux above 100 MeV will be at least 2.4 $\times
10^{-9}$ photons cm$^{-2}$ s$^{-1}$ can be computed:
\be \left< k\right>_{min} = \left( \frac{M_{H_2}}{10^9 {\rm
M}_\odot} \right)^{-1} \left( \frac{D_L}{\rm Mpc} \right)^2 \,.
\label{kmin} \ee
This approach represents a first step in order to establish the
plausibility of the future detection of a given LIRG or SG in the
$\gamma$-ray band. A similar estimation can be made for the TeV
flux expected from these objects. V\"olk et al. (1996) found:
\be F(>1 \; {\rm TeV}) \sim 1.7 \times 10^{-13} \left(
\frac{E}{\rm TeV} \right)^{-1.1} \left( \frac{M}{10^9 {\rm
M}_\odot} \right) \left( \frac{D_L}{\rm Mpc} \right)^{-2} k \;\;\;
{\rm photons\; cm^{-2} s^{-1}} \,, \ee
where a power law slope of 2.1 is assumed for the CR spectrum.
{For comparison, it is useful to keep in mind that an} integrated
flux above 1 TeV of about $1\times 10^{-13}$ photons cm$^{-2}$
s$^{-1}$ is the expected 5$\sigma$ flux sensitivity for a 50 hr
observation at small zenith angle of the new ground-based imaging
atmospheric \v{C}erenkov telescopes (IACTs). Then, those galaxies
that might appear in the new GeV catalogs might also constitute
new targets for the ground-based telescopes at higher energies,
provided their proton spectrum are sufficiently hard.\footnote{ In
addition, the signal-to-noise ratio in neutrino telescopes
(neutrinos will be unavoidably produced in hadronic interactions
leading to charged pions) can be approximately computed starting
from the $\gamma$-ray flux (see, e.g., Anchordoqui et al. 2003b).
LIRGs could be new candidate sources for ICECUBE if they are
detectable sources of TeV photons.}

\begin{figure*}[t]
\centering
\includegraphics[width=4cm]{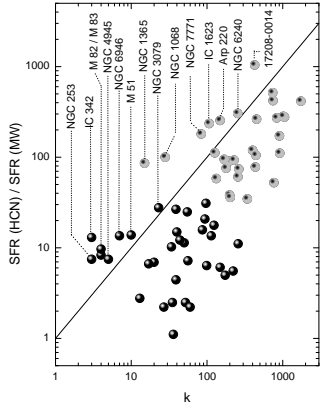}
\caption{Plausible values of enhancements for the HCN galaxies
obtained as the ratio between the SFR of each galaxy and that of
the Milky Way versus the needed one for them to be detectable by
GLAST. LIRGs and starbursts (less luminous IR galaxies) are shown
as white points (black) points. From Torres et al. (2004).}
\label{kvskplau}
\end{figure*}

However, obtaining reasonable values for the minimum $\langle
k\rangle$ required for detection is not {a sufficient} condition
to claim the plausibility for a galaxy to be a gamma-ray source:
it is also needed that the particular galactic environment is
active enough as to provide the minimum computed $\langle k
\rangle$ value.
A first indication of the average value of the CR enhancement in a
given galaxy can be indirectly estimated from the star formation
rate (SFR), quantity that can be directly related to observations.
{ More star formation implies a higher supernova explosion rate,
and as this happens, there are more cosmic ray accelaration sites
and thus an enhanced cosmic ray density.} Therefore, a reasonable
first assumption is (e.g., Drury et al. 1994, Aharonian \& Atoyan
1996, Torres et al. 2003, etc.):
\be \langle k \rangle  \equiv  \frac{q_\gamma}{q_{\gamma,\oplus}}
\sim \frac{\omega_{CR}}{\omega_{CR,\oplus}}  \sim  \frac{SN
rate}{SN rate_{\rm MW}}  \sim  \frac{SFR}{SFR_{\rm MW}} \,. \ee
The SFR is highly correlated to the quantity of dense molecular
gas present in the galaxy, as can be seen in Gao and Solomon
(2003b) figure 6,
\be SFR = 1.8 \left( \frac{M_{dense}}{10^8 M_{\odot}} \right)
\left( \frac{10}{\alpha_{\rm HCN}} \right) \;\; {\rm M}_{\odot}
{\rm yr}^{-1} \,. \ee
The dense mass is traced by the HCN emission and it is found to be
proportional to the HCN luminosity (Gao and Solomon 2003a):
\be M_{dense} = \alpha_{\rm HCN} L_{\rm HCN} \sim 10 \left(
\frac{L_{\rm HCN}}{\rm K \; km \; s^{-1} \, pc^2} \right) \;\;
{\rm M}_{\odot} \,. \ee
Therefore:
\be SFR = 18 \left( \frac{L_{\rm HCN}} {10^8 \, {\rm K \; km \;
s^{-1} \, pc^2}} \right) \;\; {\rm M}_{\odot} {\rm yr}^{-1} \,.
\label{sfr} \ee
The Milky Way star formation rate (quoted as $SFR_{\oplus}$) can
be also estimated from Equation (\ref{sfr}), being the HCN
luminosity $L_{\rm HCN}(MW) \sim$ 0.04 $\times 10^8 \, {\rm K \;
km \; s^{-1} \, pc^2}$ (e.g., Solomon et al. 1992, Wild \& Eckart
2000). Then, a plausible value of the CR enhancement (obtained as
the ratio between the SFR of the galaxy and that of our Milky Way)
can be computed for each HCN galaxy. Figure \ref{kvskplau} shows
these values versus the needed $k$ in order to make the galaxy
detectable by GLAST/LAT. Only galaxies appearing above or around
the line of unit slope can be considered prime candidates for
detection.
While a galaxy with high $L_{\rm HCN} / L_{\rm CO}$  ratio (i.e.,
with a high mass fraction of dense gas) will be a LIRG (or a
ULIRG), the converse is not always true (Gao \& Solomon 2003a).
There are gas-rich galaxies which are LIRGs only because of the
huge amount of molecular gas they possess, not because they have
most of it at high density (and thus are undergoing a particularly
strong starburst phenomenon). In some of these cases, while the
value of enhancement needed for detection might only be of  a few
hundreds, the plausible value of $k$ is much lower, since no
strong star formation is ongoing (e.g., NGC 1144, Mrk 1027, NGC
6701, and Arp 55, which appear to be using the huge molecular mass
they have in creating stars at a normal SFR). In the context of
$\gamma$-ray observability, GLAST will detect those galaxies that,
being close enough, not only shine in the FIR but that do so {\it
because} of their active strong star formation processes.

\section{More detailed models: how to?}

\begin{figure*}[t]
\centering
\includegraphics[width=0.8\textwidth]{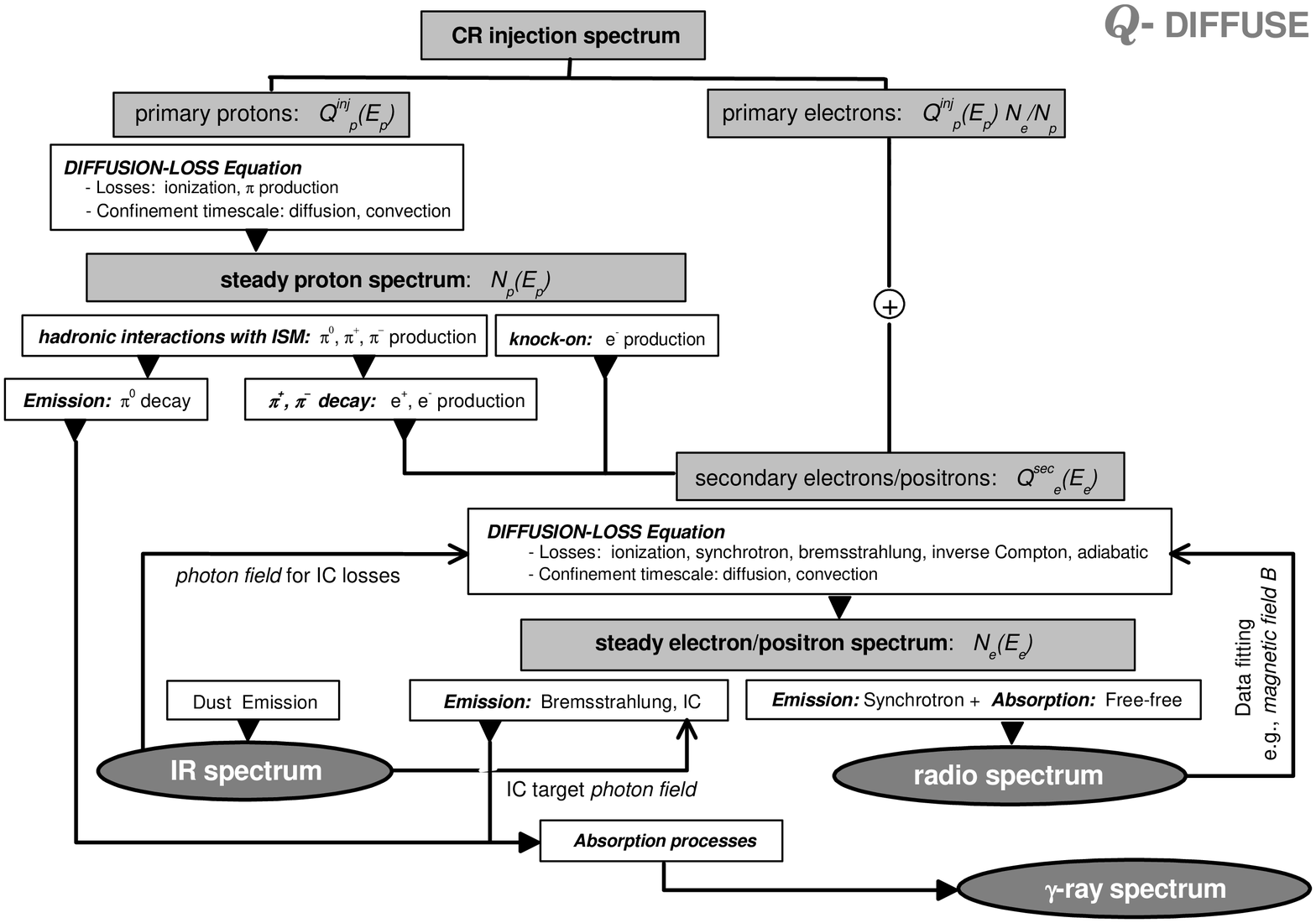}
\caption{${\cal Q}$-{\sc diffuse} flow diagram.} \label{flow}
\end{figure*}

More detailed theoretical predictions require numerical modelling.
An example of such approach is provided by  the numerical package
${\cal Q}$-{\sc diffuse} (Torres 2004). The logic of the code is
shown schematically in Figure \ref{flow}. As starting point, an
injection spectrum of primary cosmic rays is assumed. Its
characteristics are estimated by linking the cosmic rays to the
known (local to the source) acceleration processes that may have
created them (e.g., supernova remnants induced shocks or
collective effects of stellar winds), and thus using the related
observational inputs, as the supernova explosion rate. The primary
proton population is subject to energy losses (by ionization and
pion production) and to escaping out from the emission region
through diffusion or convection processes. From the proton steady
state (i.e. the solution of the diffusion-loss equation), the
computation of the secondaries is done. Secondary electrons and
positrons are generated from knock-on interactions and decay of
secondary charged pions. The contribution of secondary electrons
and positrons is then summed up to the primary electron
population, which is generally assumed to be the same than the
proton injection spectrum but by a factor $N_e/N_p$. The whole
electron/positron population is then let to evolve to its steady
state, computing the energy losses that it is subject to
(synchrotron emission, ionization, bremsstrahlung, inverse
Compton, and adiabatic losses) and the confinement timescale. The
radio spectrum is then evaluated from the steady electrons
synchrotron emission, modulated by free-free absorption of the
medium plasma electrons. A comparison between the predicted radio
flux and the existing data provides a direct feedback to some
model parameters that appear in the computation of the electron
steady state. They are tuned until a good agreement with the
measured radio spectrum is achieved. For example, one of the
crucial parameters, which indeed is hardly known with good
precision in most of the scenarios but plays an important role
both in the electron synchrotron energy losses and in the
synchrotron radio emission, is the magnetic field. { Other
parameter crucial in explaining low frequency radio data is the
absorption frequency of free-free radiation (alternatively, the
emission measure and plasma temperature). These two, are
determined within the code until an agreement with observations is
reached.} Infrared emission from dust existing in the region is
also simulated with the parameters that characterize the infrared
photon field being adjusted to describe the observational existing
data. These photons are the seed of inverse Compton process,
either when considering the electron energy losses or the
resultant high energy $\gamma$-ray emission. Once all the
parameters have been fixed and the proton and electron steady
populations have been determined, high energy $\gamma$-ray
emission is evaluated. This includes the decay of neutral pions,
and leptonically-generated $\gamma$-rays, through bremsstrahlung
and inverse Compton of the steady electron population. Photon
absorption through gamma-gamma and gamma-Z processes are
considered to obtain the final predictions of fluxes.

\section{Example: the case of NGC 253}

\subsection{Phenomenology of the central region of NGC~253}

NGC~253 is located at a distance of $\sim 2.5$ Mpc and it is a
nearly edge-on (inclination ~78$^o$) barred Sc galaxy. The
continuum spectrum of NGC~253 has a luminosity of $4 \times
10^{10}$ L$_\odot$ (Melo et al. 2002). The FIR luminosity is at
least a factor of 2 larger than that of our own Galaxy, and it
mainly comes from the central nucleus. IR emission can be
understood as cold ($T \sim 50 K$) dust reprocessing of stellar
photon fields.

%
When observed at 1 pc resolution, at least 64 individual compact
radio sources have been detected within the central 200 pc of the
galaxy (Ulvestad \& Antonucci 1997), and roughly 15 of them are
within the central arcsec of the strongest radio source,
considered to be either a buried active nucleus or a very compact
SNR. Of the strongest 17 sources, about half have flat spectra and
half have steep spectra. This indicates that perhaps half of the
individual radio sources are dominated by thermal emission from H
II regions, and half are optically thin synchrotron sources,
presumably SNRs. There is no compelling evidence for any sort of
variability in any of the compact sources over an 8 yr time
baseline.
The region surrounding the central ~200 pc has also been observed
with subarcsec resolution and 22 additional radio sources stronger
than 0.4 mJy were detected within 2kpc of the galaxy nucleus
(Ulvestad 2000). The region outside the central starburst may
account for about 20\% of the star formation of NGC~253. It is
subject to a supernova explosion rate well below $0.1$ yr$^{-1}$,
and has an average gas density in the range 20--200 cm$^{-3}$,
much less than the most active nuclear region (Ulvestad 2000).

%
Carilli (1996) presented low frequency radio continuum
observations of the nucleus at high spatial resolution. Free-free
absorption was claimed to be the mechanism producing a flattening
of the synchrotron curve at low energies, with a turnover
frequency located between 10$^{8.5}$ and 10$^9$ Hz. The emission
measures needed for this turnover to happen, for temperatures in
the order of 10$^4$ K, is at least 10$^5$ pc cm$^{-6}$.
%
%
As shown by infrared, millimeter, and centimeter observations, the
200 pc central region dominates the current star formation in
NGC~253, and is considered as the starburst central nucleus (e.g.,
Ulvestad and Antonucci 1997, Ulvestad 2000). Centimeter imaging of
this inner starburst, and the limits on variability of radio
sources, indicates a supernova rate less than 0.3 yr$^{-1}$
(Ulvestad \& Antonucci 1997), which is consistent with results
ranging from 0.1 to 0.3 yr$^{-1}$ inferred from models of the
infrared emission of the entire galaxy (Rieke et al. 1980; Rieke,
Lebofsky \& Walker 1988, Forbes et al. 1993). When compared with
Local Group Galaxies, the supernova rate in NGC~253 is one order
of magnitude larger 
(Pavlidou and Fields 2001).

Current estimates of the gas mass in the central 20'' -- 50'' ($<
600$ pc) region range from 2.5 $\times 10^{7}$ M$_\odot$
(Harrison, Henkel \& Russell 1999) to 4.8 $\times 10^{8}$
M$_\odot$ (Houghton et al. 1997), see Bradford et al. (2003),
Sorai et al. (2000), and Engelbracht et al. (1998) for
discussions. For example, using the standard CO to gas mass
conversion, the central molecular mass was estimated as 1.8
$\times 10^{8}$ M$_\odot$ (Mauersberger et al. 1996). It would be
factor of $\sim 3$ lower if such is the correction to the
conversion factor in starburst regions which are better described
as a filled intercloud medium, as in the case of ULIRGs, instead
of a collection of separate large molecular clouds, see Solomon et
al. (1997), Downes \& Solomon (1998), and Bryant \& Scoville
(1999) for discussions. Thus we will assume in agreement with the
mentioned measurements that within the central 200 pc, a disk of
70 pc height has $\sim$ 2 $\times 10^{7}$ M$_\odot$ uniformly
distributed, with a density of $\sim 600$ cm$^{-3}$. Additional
target gas mass with an average density of $\sim$50 cm$^{-3}$ is
assumed to populate the central kpc outside the innermost region,
but subject to a smaller supernova explosion rate $\sim 0.01$
yr$^{-1}$, 10\% of that found in the most powerful nucleus
(Ulvestad 2000).

\subsection{The multi-frequency emission from NGC 253}

\begin{figure*}[t]
\centering
\includegraphics[width=4cm,height=5cm]{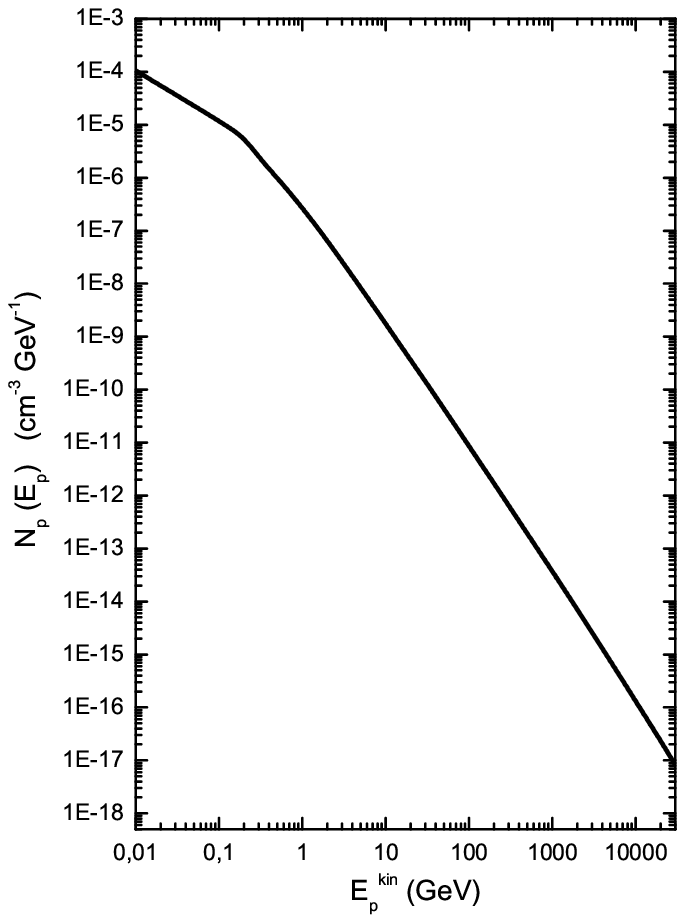}
\includegraphics[width=4cm,height=5cm]{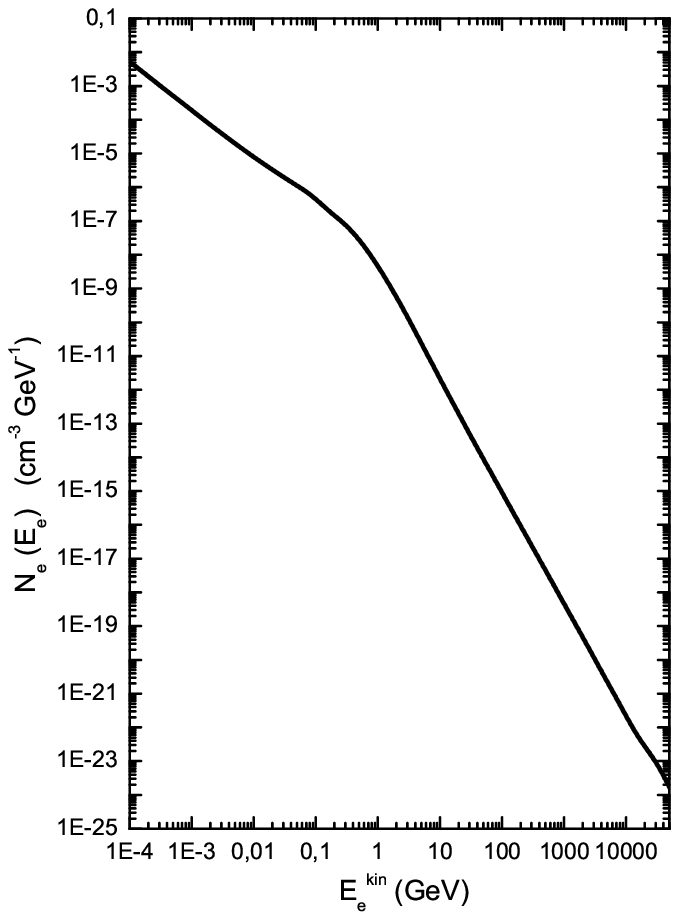}
\includegraphics[width=4cm,height=5cm]{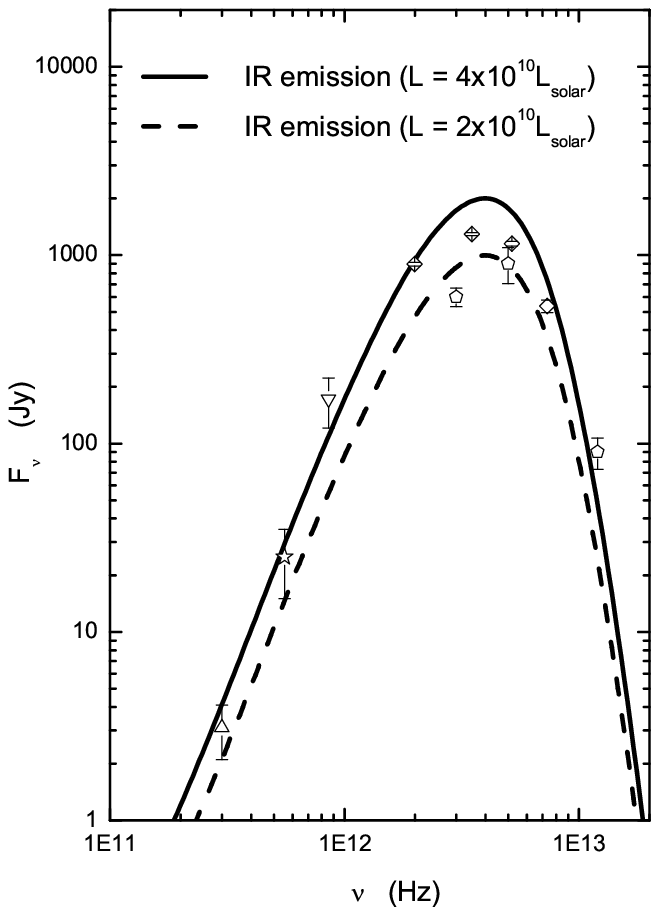}
\includegraphics[width=4cm,height=5cm]{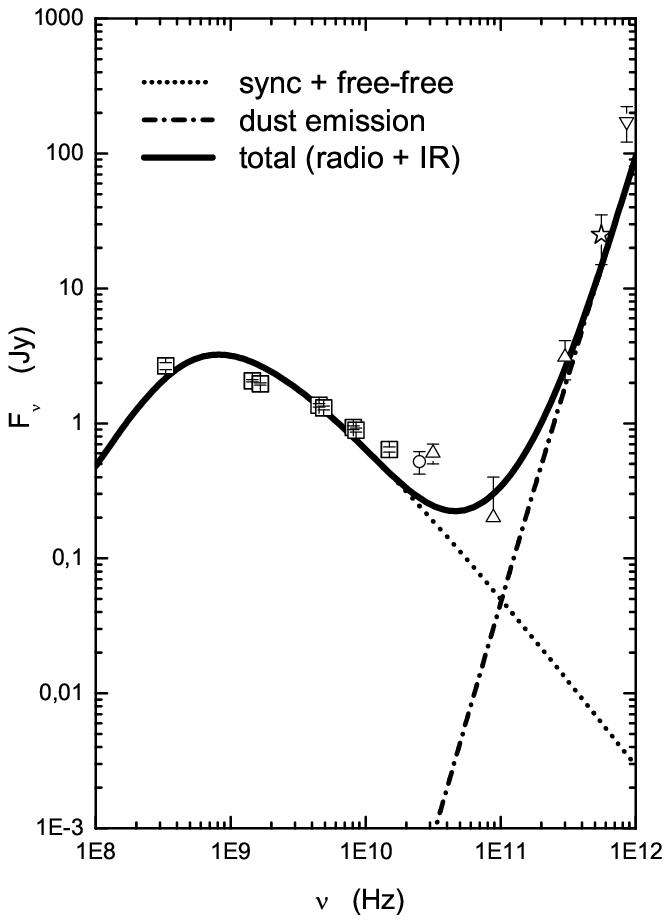}
\includegraphics[width=4cm,height=5cm]{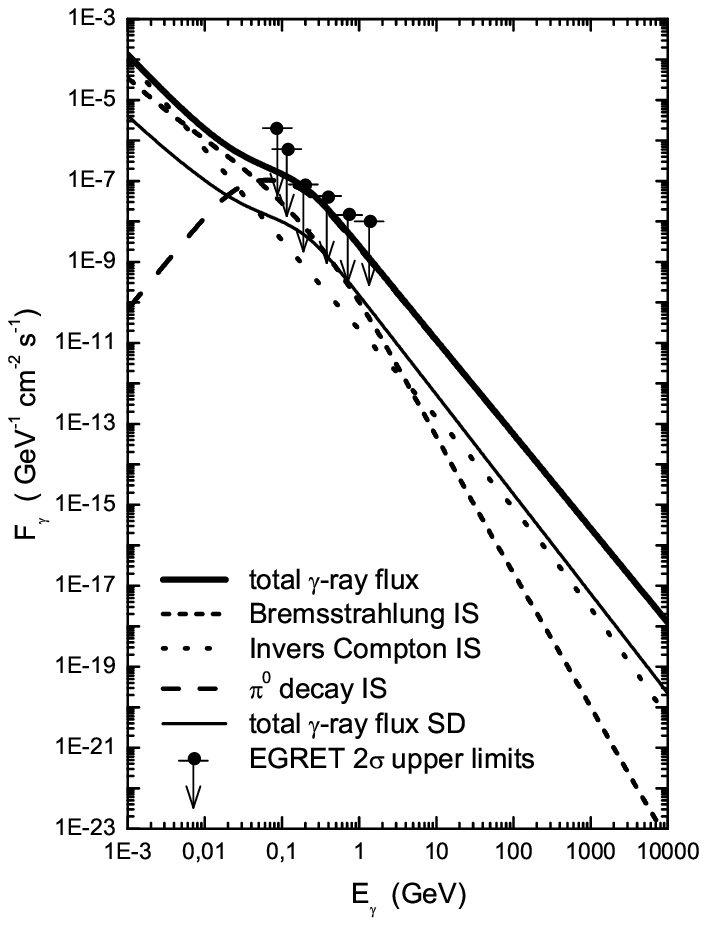}
\includegraphics[width=4cm,height=5cm]{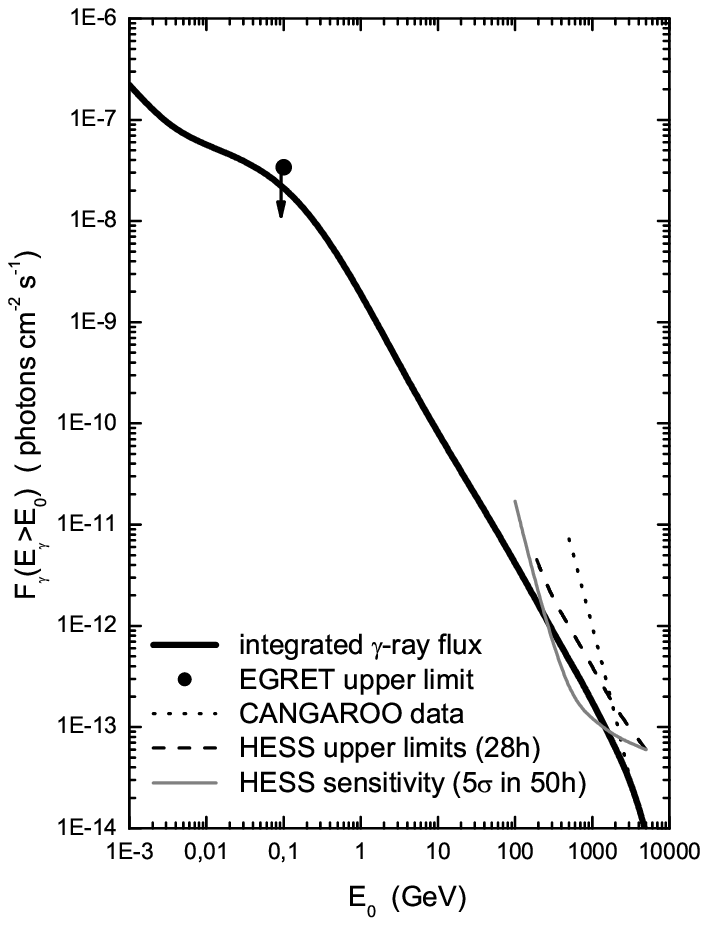}
\caption{From top to bottom and left to right: a) Steady proton
distributions in the innermost region of NGC~253. b) Idem for the
steady electron distribution. c) IR flux from NGC~253 assuming a
dilute blackbody with temperature $T_{\rm dust }= 50$ K and
different total luminosities. d) Multifrequency spectrum of
NGC~253 from radio to IR and comparison with experimental data
points. e) Differential $\gamma$-ray fluxes from the central
region of NGC~253. Total contribution of the surrounding disk is
separately shown, as are the EGRET upper limits. Also shown are
the relative contributions of bremsstrahlung, inverse Compton, and
neutral pion decay to the $\gamma$-ray flux. f) Integral
$\gamma$-ray fluxes. The EGRET upper limit (for energies above 100
MeV), the CANGAROO  (Itoh et al. 2002) integral flux as estimated
from their fit, the (4 telescopes) HESS sensitivity (for a
5$\sigma$ detection in 50 hours), and the (2 telescopes) HESS
upper limit curve on NGC~253 (Aharonian et al. 2005) are given.
Absorption effects are already taken into account. From
Domingo-Santamar\'ia and Torres (2005).} \label{steady}
\end{figure*}

The numerical solution of the diffusion-loss equation for protons
and electrons in NGC 253 is shown in Figure \ref{steady}a and
\ref{steady}b for an adopted  diffusive residence timescale of 10
Myr, a convective timescale of 1 Myr, and a density of $\sim 600$
cm$^{-3}$. In the case of electrons, the magnetic field with which
synchrotron losses are computed in Figure \ref{steady}b is 300
$\mu$G. The latter is fixed requiring that the steady electron
population produces a flux level of radio emission matching
observations. An injection electron spectrum is considered --in
addition to the secondaries-- in generating the steady electron
distribution. From about $E_e-m_e \sim 10^{-1}$ to $10$ GeV, the
secondary population of electrons dominates, in any case. The IR
continuum emission is modelled with a spectrum having a dilute
blackbody (graybody) emissivity law, proportional to
$\nu^{\sigma}B(\epsilon, T)$, where $B$ is the Planck function.
Figure 1c shows the result of this modelling and its agreement
with observational data when the dust emissivity index
$\sigma=1.5$ and the dust temperature $T_{\rm dust}= 50 K$.
Whereas free-free emission is subdominant when compared with the
synchrotron flux density, free-free absorption plays a key role at
low frequencies, determining the opacity. We have found a
reasonable agreement (see Figure 1d) with all observational data
for a magnetic field in the innermost region of 300 $\mu$G, an
ionized gas temperature of about 10$^4$ K, and an emission measure
of $5 \times 10^5$ pc cm$^{-6}$. Figure 1e shows bremsstrahlung,
inverse Compton, and pion decay $\gamma$-ray fluxes from the
central nucleus of NGC 253. These results are obtained with the
model in agreement with radio and IR observations. These
predictions, while complying with EGRET upper limits, are barely
below them. If this model is correct, NGC~253 is bound to be a
bright $\gamma$-ray source for GLAST. The integral fluxes are
shown in Figure 1f. This model predicts that, given enough
observation time, NGC~253 is also to appear as a point-like source
in an instrument like HESS. {The HESS array presented   results
for NGC 253, based on a total of 28 hours taken during the
construction of the array with 2 telescopes operating (Aharonian
et al. 2005). The energy threshold for this dataset was 190 GeV.
Upper limits from HESS on the integral (99 \% confidence level)
are shown in Figure 1f. As an example, above 300 GeV, the upper
limit is $1.9 \times 10^{-12}$ photons cm$^{-2}$ s$^{-1}$. The
predictions for NGC 253 are below these upper limits at all
energies but still above HESS sensitivity for reasonable
observation times.}  The effect of the opacity on the integral
$\gamma$-ray fluxes only plays a role above 3 TeV.

\section{Predictions at the highest energies and
differential cross section parameterization: application to Arp 220}

Numerical calculations of the $\gamma$-ray flux produced through
$\pi^0$ decay involve the differential cross-section of
proton-proton (pp) interactions, for which  experimental data to
confront with only exists up to the GeV range. Therefore,
extrapolation of the differential pp cross section to higher
energies (as needed in the kind of scenarios herein discussed)
induces large uncertainties in the resultant $\gamma$-ray flux. An
analysis of the effect of the currently available differential
cross section parameterizations over the high energy $\gamma$-ray
flux can be found in the appendix of Domingo-Santamar\'ia and
Torres (2005).
%
For $\gamma$-rays above few TeV (i.e., $\gamma$-rays mostly
generated by protons above few tens of TeV), the Blattnig et al.
(2000) differential cross section parameterization makes the
$\gamma$-ray emitted spectrum much harder than the proton spectrum
that produced them. This indicates that a direct extrapolation of
Blattnig et al. parameterization above TeV induces significant
overpredictions of fluxes.

\begin{figure*}[t]
\centering
\includegraphics[width=.45\textwidth]{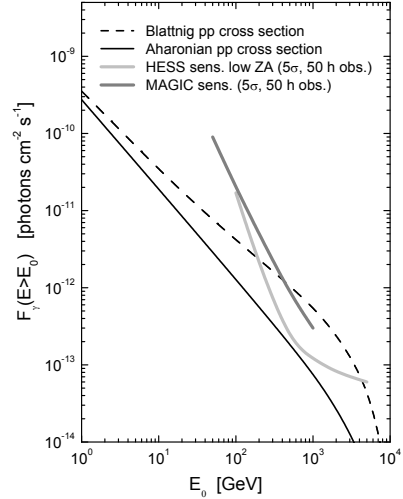}
\caption{Total integral flux predictions for Arp 220. The dashed
line shows the results obtained with the ${\cal Q}$-{\sc diffuse}
numerical package using Blattnig et al. (2000) parameterization of
the pp cross section; the solid line shows the fluxes obtained
with the same model but using other cross sections (Aharonian \&
Atoyan, or Kamae et al.). The HESS and MAGIC telescopes
sensitivities, 50 hours of observation time for a 5$\sigma$
detection, for low zenith angles (although note that this is shown
here just for quick comparison, since HESS can only observed Arp
220 above $\sim$ 50$^\circ$), are also shown to remark the
differences in predictions for observability that the use of one
or other cross section can induce. Absorption effects are already
taken into account in both cases.} \label{g-emisarp}
\end{figure*}

Consider Arp 220, for which there is a multiwavelength model
already presented in the literature (Torres 2004).\footnote{Arp
220 is the nearest ultra-luminous infrared galaxy, in fact it is
the only one inside the 100 Mpc sphere, and the best studied.
} At the time this modelling was made, it was unclear whether the
Blattnig et al. (2000) parameterization (which actually produces
the right total cross section at all energies) would also produce
the correct differential cross section above 100 GeV, and it was
then incorporated into the ${\cal Q}$-{\sc diffuse} code to
compute the neutral pion $\gamma$-ray production. As discussed,
this yield to overestimates of fluxes above 100 GeV, where
Blattnig et al. parameterization should not be used. New estimates
of the neutral pion decay contribution to the $\gamma$-ray flux
expected from Arp 220 have been calculated for this paper using
more reliable approaches to the cross section at high energies,
and are shown in Figure \ref{g-emisarp}, see Domingo-Santamar\'ia
and Torres (2005) for a discussion on the parameterizations used.
This comparison uses the same galactic model in every other
respect, but just different differential cross sections.

Although the prediction of a possible detection of Arp 220 by the
GLAST satellite is still supported (all parameterizations provide
the same result at lower energies), its detection in the IACTs
regime now looks much more difficult. The total predicted fluxes
in $\gamma$-rays above 300 GeV and 1 TeV are $\sim 3 \times
10^{-13}$ photons cm$^{-2}$ s$^{-1}$ and $\sim 8 \times 10^{-14}$
photons cm$^{-2}$ s$^{-1}$, respectively. It can be seen in Figure
\ref{g-emisarp} that Arp 220 is barely below the HESS sensitivity
curve (who has to observe Arp 220 at large zenith angle because of
its location) and also below MAGIC capabilities unless very deep
observations are performed.

\section{Concluding remarks}

Nearby galaxies, and especially those that are starbursts and
luminous infrared galaxies, and thus are subject to a high level
of star formation, are prime candidates to constitute a new
population of high energy gamma-ray emitters. The definitive
observational proof of this statement will most likely come with
the launch of GLAST and with further observations of ground
\v{C}erenkov telescopes.

\section*{Acknowledgments}

The work of DFT was performed under the auspices of the U.S.
D.O.E. (NNSA), by the University of California Lawrence Livermore
National Laboratory under contract No. W-7405-Eng-48. The work of
ED-S was done under a FPI grant of the Ministry of Science and
Technology of Spain.

\end{document}